\documentclass[aps,twocolumn,showpacs,preprintnumbers,nofootinbib,prd,10pt,superscriptaddress]{revtex4-1}

\makeatletter
\def\l@subsubsection#1#2{}
\def\l@subsubsubsection#1#2{}
\makeatother

\usepackage{graphicx,amssymb,amsmath,amsthm,amsfonts,epsfig,epsf}
\usepackage{comment}
\usepackage{mathrsfs}
\usepackage[usenames]{color}
\usepackage{epstopdf}
\usepackage{bm}
\usepackage{dcolumn}
\usepackage{latexsym}
\usepackage{rotating}
\usepackage{longtable}

\usepackage{enumerate}
\usepackage{tensor,multirow}
\usepackage{url}
\usepackage[dvipsnames]{xcolor}
\usepackage[unicode]{hyperref}
\hypersetup{colorlinks=true, citecolor=MidnightBlue,
            linkcolor=MidnightBlue, urlcolor=MidnightBlue, linktocpage=true}
\usepackage{orcidlink}

\newcommand{\dd}{\mathrm{d}}
\newcommand{\MBH}{M_{\rm BH}}

\newcommand{\CSC}{\mathcal{C}_{\rm sc}}
\renewcommand{\Re}{{\rm{Re}}}
\renewcommand{\Im}{{\rm{Im}}}

\begin{document}

\title{Gravitational Atom Spectroscopy}
\author{Matteo Della Rocca~\orcidlink{0009-0001-4470-3694}}
\email{matteo.dellarocca@phd.unipi.it}
\affiliation{Dipartimento di Fisica, Universit\`a di Pisa, Largo B. Pontecorvo 3, 56127 Pisa, Italy}
\affiliation{INFN, Sezione di Pisa, Largo B. Pontecorvo 3, 56127 Pisa, Italy}
\author{Thomas F.M.~Spieksma~\orcidlink{0000-0003-0513-5580}}
\affiliation{Center of Gravity, Niels Bohr Institute, Blegdamsvej 17, 2100 Copenhagen, Denmark}
\affiliation{Rudolf Peierls Centre for Theoretical Physics, University of Oxford, Parks Road, Oxford OX1 3PU, United Kingdom}
\author{Francisco Duque~\orcidlink{0000-0003-0743-6491}}
\affiliation{Max Planck Institute for Gravitational Physics (Albert Einstein Institute) 
Am Mühlenberg 1, D-14476 Potsdam, Germany}
\author{Leonardo Gualtieri~\orcidlink{0000-0002-1097-3266}}
\affiliation{Dipartimento di Fisica, Universit\`a di Pisa, Largo B. Pontecorvo 3, 56127 Pisa, Italy}
\affiliation{INFN, Sezione di Pisa, Largo B. Pontecorvo 3, 56127 Pisa, Italy}
\author{Vitor Cardoso~\orcidlink{0000-0003-0553-0433}}
\affiliation{Center of Gravity, Niels Bohr Institute, Blegdamsvej 17, 2100 Copenhagen, Denmark}
\affiliation{CENTRA, Departamento de F\'{\i}sica, Instituto Superior T\'ecnico -- IST, Universidade de Lisboa -- UL,
Avenida Rovisco Pais 1, 1049 Lisboa, Portugal}
\begin{abstract}
Black holes in our Universe are rarely truly isolated, being instead embedded in astrophysical environments such as plasma or dark matter. A particularly intriguing possibility is that light scalar fields form bound states around black holes, producing extended ``clouds'' known as \textit{gravitational atoms}. When these clouds become sufficiently compact, the spacetime can no longer be described by a vacuum solution of General Relativity. In this regime, one can construct quasi-stationary, spherically symmetric, self-gravitating scalar gravitational-atom configurations. Here, we explore an observationally relevant aspect of these systems by computing their fundamental quasi-normal mode. We present a fully relativistic calculation of the axial modes in both the time and frequency domains, finding frequency shifts relative to the vacuum case that depend mostly on the compactness of the gravitational atom. For sufficiently compact configurations, these shifts may be detectable by current or future gravitational wave detectors. 
\end{abstract}
\maketitle
\section{Introduction}
Gravitational waves (GWs) emitted after the merger of two compact objects offer a direct way to probe the properties of the remnant. According to black hole (BH) perturbation theory, the post-merger signal enters a \textit{ringdown} phase, during which the emitted GWs can be described as a superposition of exponentially damped sinusoids whose frequencies correspond to the quasi-normal modes (QNMs) of the remnant. The ringdown of a BH depends only on its macroscopic parameters~\cite{Kokkotas:1999bd,Berti:2009kk,Ferrari:2007dd,Konoplya:2011qq,Berti:2025hly}, which are the mass, the spin, and, if present, the charge, as dictated by the uniqueness theorems of General Relativity (GR)~\cite{Israel:1967wq,Carter:1971zc,Robinson:1975bv}. The study of this relaxation stage, known as \textit{BH spectroscopy}, provides a powerful means to characterise the merger remnant and serves as an independent consistency check of inspiral-based analyses.
This picture, however, assumes that the merger takes place in vacuum. While recent work has shown that typical galactic-scale environments, such as dark matter halos, have negligible impact on the ringdown at the sensitivity of next-generation detectors~\cite{Spieksma:2024voy}, much denser or more compact environments could leave detectable imprints. Indeed, several toy models predict sizeable deviations when the environment is sufficiently dense or compact~\cite{Cardoso:2020nst,Cardoso:2021wlq,Cardoso:2022whc,Cannizzaro:2024yee,Pezzella:2024tkf}. 
A particularly interesting class of such environments involves {\it ultralight scalar fields} with masses in the range $\sim\!\!\left[10^{-22}-\!10^{-11}\right]\mathrm{eV}$. These fields arise naturally in~many extensions of the Standard Model and can, for example, address the strong-CP problem~\cite{Peccei:1977hh, Weinberg:1977ma,Wilczek:1977pj} or emerge from string theory~\cite{Svrcek:2006yi,Arvanitaki:2009fg}.~They have also been extensively explored in cosmology as dark-matter candidates~\cite{Bergstrom:2009ib,Marsh:2015xka,Hui:2016ltb,Ferreira:2020fam, Hui:2021tkt}. Their extremely small masses and weak couplings make them inaccessible to conventional particle experiments, yet gravity offers a unique observational window. In particular, their large Compton wavelength allows them to gravitationally couple with astrophysical BHs. 
Although stationary scalar ``hair'' is excluded by no-hair theorems~\cite{Bekenstein:1995un,Cardoso:2016ryw}, BHs can support time-dependent, long-lived scalar configurations, commonly referred to as \textit{scalar wigs} or \textit{gravitational atoms}~\cite{Barranco:2011eyw,Baumann:2019eav,Cardoso:2022vpj,Zhong:2023xll,Alcubierre:2024mtq}. These structures may form through accretion of ambient scalar fields in galactic centres~\cite{Hui:2019aqm,Clough:2019jpm,Cardoso:2022nzc}, via dark matter relaxation~\cite{Budker:2023sex} or through superradiant instabilities (see~\cite{Brito:2015oca,Herdeiro:2015waa} and references therein). Depending on the scalar mass, gravitational atoms can develop around BHs spanning the entire mass spectrum, from stellar to supermassive, potentially affecting the GW signal emitted during binary coalescences at all scales~\cite{Bamber:2022pbs,Aurrekoetxea:2023jwk,Tomaselli:2023ysb,Duque:2023seg,Tomaselli:2024bdd,Aurrekoetxea:2024cqd,Tomaselli:2024dbw,DellaMonica:2025zby,Dyson:2025dlj,Roy:2025qaa,Cheng:2025wac}.
In this work, we extend the BH spectroscopy programme to include gravitational atoms. We focus on the axial sector and quantify how the presence of the scalar field modifies the frequency and damping time of the system's fundamental QNM. Our analysis is performed in both the time and frequency domains, finding excellent agreement between the two. We further show that existing LVK observations could, in principle, already be sensitive to the deviations induced by gravitational atoms in certain regions of parameter space.
Throughout this work, we adopt geometric units ($G=c=1$) and denote the scalar-field mass by $\mu_{\rm s}\hbar$, such that its reduced Compton wavelength is $1/\mu_{\rm s}$.
\section{Gravitational atoms}\label{sec:scalar_wigs}
We consider a massive, complex scalar field minimally coupled to gravity, described by the action
\begin{equation}
 S\left[\boldsymbol{g}, \Phi\right]= \int \mathrm{d}^4 x \frac{\sqrt{-g}}{16 \pi} \left(R - \frac{1}{2} \nabla_\mu \Phi \nabla^{\mu} \Phi^* - \frac{1}{2} \mu_{\rm s}^2 |\Phi|^2 \right)\, ,
 \label{eq:action}
 \end{equation}
where $R$ is the Ricci scalar, $g$ the determinant of the metric $\boldsymbol{g}$, and $\mu_{\rm s}\hbar$ the mass of the scalar field. Varying the action yields the Einstein-Klein-Gordon equations,
\begin{equation}
    G_{\mu\nu}= T_{\mu\nu}\,, \qquad \left(\nabla_\mu\nabla^\mu -\mu_{\rm s}^2\right)\Phi=0\,, \label{eq:ekg_eqs}
\end{equation}
where $G_{\mu\nu}$ is the Einstein tensor, and the stress-energy tensor is given by
\begin{equation}
    T_{\alpha\beta}= 2 \nabla_{(\alpha} \Phi \nabla_{\beta)} \Phi^*-g_{\alpha \beta}\left(\nabla_\sigma \Phi \nabla^\sigma \Phi^*+\mu_{\rm s}^2 \Phi \Phi^*\right)\,.
\label{eq:T}
\end{equation}
Solutions of the Klein–Gordon equation~\eqref{eq:ekg_eqs} on a Kerr background admit {\it quasi-bound states}:~test-field configurations that are purely ingoing at the horizon and exponentially suppressed at spatial infinity~\cite{Detweiler:1980uk,Dolan:2007mj,Barranco:2011eyw}. Linear combinations of such states are commonly referred to as {\it gravitational atoms}~\cite{Barranco:2012qs,Baumann:2019eav}. When the backreaction of the scalar field on the metric is included -- i.e., when the coupled Einstein–Klein–Gordon equations~\eqref{eq:ekg_eqs} are solved self-consistently -- the system forms a {\it self-gravitating} gravitational atom~\cite{Barranco:2017aes,Cardoso:2022vpj,Zhong:2023xll,Alcubierre:2024mtq}. Time-domain evolutions indicate that, after an initial transient, generic scalar-field configurations around BHs evolve towards states that are well approximated by gravitational atoms, largely independent of the initial data~\cite{Barranco:2012qs,Cardoso:2022vpj,Zhong:2023xll}.

\begin{figure}
    \centering
    \includegraphics[width=0.98\linewidth]{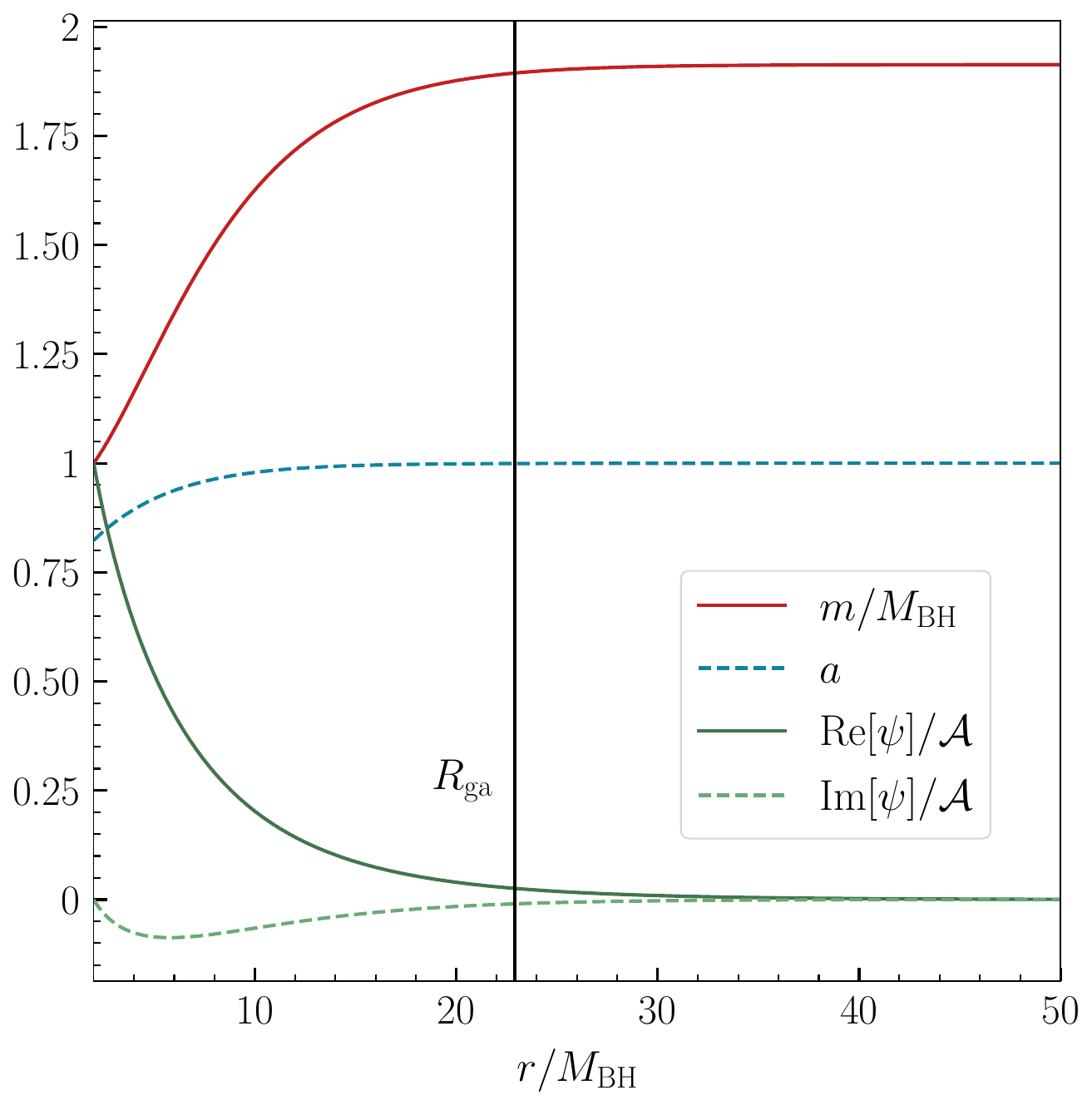}
    \caption{Background solution for $\alpha = 0.249$ and $\mathcal{A} = 0.401$. Shown are the mass function $m/\MBH$ (solid red), which asymptotes to the constant value $M_{\mathrm{ga}} = 1.913 M_{\mathrm{BH}}$;~the metric function $a$ (dashed blue), which approaches unity at large radii;~and the real (solid dark green) and imaginary (dashed light green) parts of the normalised scalar field $\psi/\mathcal{A}$, both decaying to zero asymptotically. The vertical line marks the characteristic radius of the gravitational atom,  $R_{\mathrm{ga}} = 22.92 M_{\mathrm{BH}}$. The corresponding eigenfrequency is $M_{\mathrm{BH}}\Omega = 0.2161 + 0.0188i$, and the compactness is $\CSC=0.03486$.}
    \label{fig:bkg1}
\end{figure}

In this work, we restrict our analysis to spherically symmetry and study the perturbations of a self-gravitating gravitational atom. We consider a scalar field oscillating at a single frequency $\Omega$, corresponding to the fundamental quasi-bound state. The background scalar field is then given by
\begin{equation}
    \Phi_0(t,r)=\psi(r)e^{i\Omega t} \,,
    \label{eq:Phi0}
\end{equation} 
where 
\begin{equation}
    \Omega=\omega-i\sigma\,,\,\mbox{ with }\, \omega,\sigma \in \mathbb{R} \ \mbox{ and }\psi: \mathbb{R}\to\mathbb{C}\,.
\end{equation}
For a non-rotating BH, quasi-bound states are always decaying~\cite{Detweiler:1980uk,Dolan:2007mj}, i.e., $\sigma<0$. Following~\cite{Barranco:2017aes,Alcubierre:2024mtq}, we adopt generalised ingoing Eddington-Finkelstein (EF) coordinates $(t,r,\theta,\varphi)$ to ensure regularity at the horizon. In these coordinates, the background metric takes the form
\begin{equation}
     ds^2=g^0_{tt}\dd t^2+2g^0_{tr}\dd t\dd r+g^0_{rr}\dd r^2+r^2(\dd \theta^2+\sin^2\theta\dd\varphi^2)\,,
\label{eq:bkg_metric}
\end{equation}
with metric functions
\begin{equation}
\begin{aligned}
    g^0_{tt}&=-a^2(t,r)\left(1-\frac{2 m(t,r)}{r}\right)\,,\\
   g^0_{tr}&= 2 \frac{a(t,r)m(t,r)}{r}\,,\\
   g^0_{rr}&=1+\frac{2m(t,r)}{r}\,.
\end{aligned}
\label{eq:metric_functions}
\end{equation}
Here, $m(t,r)$ is the Misner-Sharp mass function and $a(t,r)$ a real, positive function. In the vacuum limit, Eq.~\eqref{eq:bkg_metric} reduces to the Schwarzschild metric in EF coordinates. Both $\psi(r)$ and the metric functions~\eqref{eq:metric_functions} are regular at the horizon, and ingoing null geodesics satisfy $a(t,r)\dd t+\dd r=\mathrm{const}$. Note that the coordinate $t$ is {\it not} the standard Schwarzschild time coordinate:~the latter diverges for geodesics crossing the horizon, whereas the EF time coordinate used here remains finite.
Substituting Eqs.~\eqref{eq:bkg_metric} and~\eqref{eq:Phi0} into Einstein's equations~\eqref{eq:ekg_eqs} yields the coupled system
\begin{equation}
    \begin{pmatrix}
    a' \\
    m'
    \end{pmatrix}=e^{2\sigma t}
    \begin{pmatrix}
    F\\
    G
    \end{pmatrix}\,,
    \label{eq:bkgEQSCompactForm}
\end{equation}
where primes denote radial derivatives. The functions $F=F(r)$ and $G=G(r)$, together with the Klein-Gordon equation for $\psi$, are given explicitly in Appendix~\ref{app:bkg}. The only explicit time dependence in the background equations arises through the factor $e^{2\sigma t}$. For quasi-stationary configurations, we restrict our analysis to timescales $T$ satisfying $2\sigma T \ll 1$. For the parameter values considered in this work, this condition holds for timescales much longer than the dynamical timescale, since $\sigma M_{\rm BH}\ll 1$ (see Fig.~\ref{fig:timedecay} in Appendix~\ref{app:bkg}). For a BH with mass $10\ M_{\odot}$ ($10^7\ M_{\odot}$) coupled to a scalar field of mass $10^{-12}\,\mathrm{eV}$ ($10^{-18}\,\mathrm{eV}$), the corresponding decay time is approximately $10^{-7}\,\mathrm{yrs}$ ($10^{-1}\,\mathrm{yrs}$). Although these timescales are larger than the dynamical timescale, they can still be shorter than the inspiral timescale. The cloud’s lifetime, however, depends sensitively on the assumption of spherical symmetry. Spinning configurations can naturally populate higher scalar field multipoles ($\ell>0$) (e.g., through superradiance~\cite{Brito:2015oca}) where an angular momentum barrier suppresses the decay of the cloud~\cite{Detweiler:1980uk}. Nevertheless, we expect the QNM shifts for the non-rotating case to be a good estimate for spinning configurations, since the imprint of the scalar field on the metric is mainly ruled by the real part of the frequency, $\omega$, and is therefore less sensitive to the spin~\cite{Detweiler:1980uk}. Within the merger dynamical timescales, the background functions can therefore be treated as stationary, i.e., $a\simeq a(r)$ and $m\simeq m(r)$.
At the horizon, $r=r_{\rm h}$, the condition $g^{0, rr}(r_{\rm h}) = 0$ implies
\begin{equation}
    r_{\rm h}=2m(r_{\rm h}) \,, 
\end{equation}
which defines the BH mass as $M_{\rm BH}=r_{\rm h}/2$. We further define
\begin{equation}
a_{\rm h}\equiv a(r_{\rm h})\,, \, \text{and} \quad 
\mathcal{A}\equiv\psi(r_{\rm h})\,,
\label{eq:bounds}
\end{equation}
where $\mathcal{A}$ can be taken as real without loss of generality, owing to the $U(1)$ symmetry of the action. The normalisation of $a_{\rm h}$ is fixed {\it a posteriori} by requiring that $a(r)\to1$ at spatial infinity.
Asymptotic flatness demands $m'\to0$ and $\psi'\to0$, which is satisfied only for a discrete set of frequencies. In what follows, we focus on the fundamental mode for which $\psi(r)$ has no nodes. The total Arnowitt–Deser–Misner (ADM) mass of the configuration is $M_{\rm ga}=m(r\to \infty)$. We define the characteristic radius of the gravitational atom, $R_{\rm ga}$, through
\begin{equation}
    m(R_{\rm ga})=99\% \ M_{\rm ga}\,,
    \label{eq:R}
\end{equation}
and the radius of the scalar cloud, $R_{\rm sc}$, defined analogously as
\begin{equation}
    m(R_{\rm sc})-\MBH=99\% \ (M_{\rm ga}-\MBH)\,.
    \label{eq:RS}
\end{equation}
The compactness of the scalar cloud is then defined as
\begin{equation}
    \CSC=\frac{M_{\rm ga}-\MBH}{R_{\rm sc}}\,.
\end{equation}
Finally, we introduce the dimensionless \emph{gravitational fine-structure constant}
\begin{equation}
    \alpha\equiv\mu_{\rm s}\MBH\,.
\end{equation}
We solve Eqs.~\eqref{eq:bkgEQSCompactForm} together with the Klein-Gordon equation for various values of $\alpha$ and ${\cal A}$. Self-gravitating gravitational atoms exist for $\alpha\le 1/4$~\cite{Barranco:2011eyw,Barranco:2013rua,Alcubierre:2024mtq}. In practice, we obtain stable numerical solutions up to $\alpha=0.249$. For $\alpha\lesssim 0.11$, the equations become stiff and the numerical scheme is unreliable. Additionally, configurations with ${\cal A}\gtrsim 0.4$ are found to be unstable against radial perturbations. An illustrative example of the background configuration for $\alpha=0.249$ and ${\cal A}=0.401$ is shown in Fig.~\ref{fig:bkg1}.
\section{Axial perturbations of a gravitational atom}\label{sec:perturbations}
Having established the background solution, we now turn to its linear perturbations. The metric is expanded as 
\begin{equation}
    g_{\mu\nu}=g^{0}_{\mu\nu}+\epsilon h_{\mu\nu}^{\rm ax}+\epsilon h^{\rm pol}_{\mu\nu}\,, 
\end{equation}
where $g^{0}_{\mu\nu}$ is the background metric~\eqref{eq:bkg_metric}. The tensors $h^{\rm ax}_{\mu\nu}$ and $h^{\rm pol}_{\mu\nu}$ denote perturbations with axial and polar parity, respectively, which are expanded in tensor spherical harmonics in the Regge-Wheeler gauge~\cite{Regge:1957td,Zerilli:1970wzz}. The scalar field is decomposed as
\begin{equation}
    \Phi(t,r,\theta,\varphi)=\Phi_0(t,r)+\epsilon\sum_{\ell m}\frac{\delta\Phi_{\ell}(t,r)}{r} Y_{\ell m}(\theta,\varphi)\,.
\label{eq:tra_S_KS}
\end{equation}
Since the scalar field is even under parity transformations and the background is spherically symmetric, scalar field perturbations couple only to the polar sector of the metric perturbations. In pursuit of a complete {\it gravitational atom spectroscopy}, we concentrate here on the simpler axial sector. Nevertheless, these modes still differ from those of a Schwarzschild BH both because the background metric~\eqref{eq:bkg_metric} is different from the Schwarzschild solution and because the perturbations couple to the background scalar field $\Phi_0$.
As in vacuum GR~\cite{Regge:1957td}, axial perturbations can be described by two functions $h^{\ell m}_{0}(t,r)$ and $h^{\ell m}_{1}(t,r)$ (see Appendix~\ref{app:ax_eqs}) whose equations reduce to a single second-order ordinary differential equation. In what follows, we omit the harmonic indices $\ell,\,m$ to avoid clutter.
\subsection{Perturbations in the frequency domain}\label{sec:ax:freq}
The QNMs of BHs are usually obtained in the frequency domain. Because QNM frequencies are complex, a direct Fourier transform of the perturbations is not well suited. Instead, we adopt the ansatz
\begin{equation}
    h_{j}(t,r)=\sum_{n} \tilde{h}_j(r,\Upsilon_n)\,e^{-i\Upsilon_n t}\,,\quad j=\{0,1\}\,,
    \label{eq:ax:f_fd}
\end{equation}
where the complex frequencies $\Upsilon_n$ are ordered by the magnitude of their imaginary part. Modes with smaller $|\Im [\Upsilon_n]|$ (lower $n$) decay slower and thus dominate the ringdown signal when the amplitude of the signal is large~\cite{Berti:2025hly}. We restrict our analysis to the fundamental, dominant mode ($n=0$), denoting its frequency by $\Upsilon\equiv\Upsilon_0$. 
The equations governing axial perturbations can be reduced to a single master equation (see Appendix~\ref{app:ax_eqs_frequency}):
\begin{equation} 
    \left(\frac{\dd^2 }{\dd r_*^2}+\Upsilon^2-V(r)\right)Q(r_*)=0\,,
\label{eq:Q}
\end{equation}
where 
\begin{equation}
    Q(r)=\frac{p(r)\tilde h_1(r)}{r}\,,
\label{eq:Q_def}
\end{equation} 
and the tortoise coordinate $r_*$ is defined by
\begin{equation}
    \frac{ \dd  r_*}{\dd r}=\frac{1}{a(1-2m/r)} \,.
    \label{eq:tortoisecoordinate}
\end{equation} 
At large radii, the scalar field vanishes and $r_*\simeq r+2M_{\rm ga}\ln(r/(2M_{\rm ga})-1)$. The potential $V(r)$ and the function $p(r)$ are given explicitly in Appendix~\ref{app:ax_eqs_frequency}, Eqs.~\eqref{eq:p}--\eqref{eq:V}. 
We compute the QNMs via a shooting method, imposing ingoing boundary conditions at the horizon and outgoing ones at infinity. Because we work in generalized EF coordinates, these conditions differ from the ``standard'' Schwarzschild case~\cite{Vishveshwara:1970zz}. At the horizon ($r = r_{\rm h}$), the metric perturbations must behave as ingoing waves,
\begin{equation}
    h_j(t,r)\propto e^{-i \Upsilon v}\,,
\end{equation}
where the ingoing null coordinate is defined as $v=t+\hat r$,\footnote{If the backreaction of the scalar cloud is neglected, $a=1$ and $v=t+r$, where $t$ corresponds to the EF time.} with
\begin{equation}
    \hat r=\int \frac{\dd r}{a(r)}\,.
\end{equation}
For radial, ingoing null geodesics, $a(r)\dd t+\dd r=0$, and therefore $v$ is constant. Substituting the above definition into Eq.~\eqref{eq:ax:f_fd} yields
\begin{equation}
    \tilde h_j(r)\sim e^{-i\Upsilon\hat r}\,.
\end{equation}
At infinity, the scalar field and the metric decouple, and the spacetime approaches a Schwarzschild solution of mass $M_{\rm ga}$. Perturbations then behave as outgoing waves,
\begin{equation}
    h_j\propto e^{-i \Upsilon u}\,,
\end{equation}
where $u=t+\hat r-2r_*$ is the outgoing null coordinate (for radial, outgoing null geodesics, $a(1-2m/r)\dd t-(1+2m/r)\dd r=0$). This implies\footnote{It is worth noting that, due to the use of generalized EF coordinates, the expansion of the perturbation function $h_j$ in Eq.~\eqref{eq:ax:f_fd} differs from the standard Regge-Wheeler derivation:~specifically, $\tilde{h}_j(r,\Upsilon_n)$ differs from the definition in~\cite{Regge:1957td} by a factor $e^{i\Upsilon(\hat r-r_*)}$.}
\begin{equation}
    \tilde h_j(r\to\infty)\sim e^{i\Upsilon (2r_*-\hat r)}\,.
\end{equation}
Since $p(r)$ (see Eq.~\eqref{eq:p}) is proportional to $e^{i\Upsilon(\hat r-r_*)}$, the corresponding boundary conditions for the master function $Q(r)$ are
\begin{equation}
\begin{aligned}
    Q&\to e^{-i \Upsilon r_*}\sum_{n=0}^{N}q_n^{\rm h}(r-r_{\rm h})^n\,,  &&r\to r_{\rm h} \,, \\
     Q&\to e^{i \Upsilon r_*}\sum_{n=0}^{M}q_n^\infty r^{-n}\,,  &&r\to \infty\,. 
\end{aligned}
\label{eq:ax:bounds}
\end{equation}
Solving the Einstein equations perturbatively at the horizon (infinity) uniquely determines the coefficients $q_{n}^{\rm h}$ ($q_{n}^\infty$) for $n>0$ in terms of $q_0^{\rm h}$ ($q_0^{\infty}$). To properly compute $q_0^{\rm h}$, we expand the background functions up to $\mathcal{O}((r-r_{\rm h})^3)$ (see Eqs.~\eqref{eq:bkg_exp_hor}). Imposing the boundary conditions~\eqref{eq:ax:bounds} then allows us to obtain the QNMs of the self-gravitating gravitational atom, as we will see in Section~\ref{sec:results}.
\subsection{Perturbations in the time domain}\label{sec:ax:time}
In the time domain, the axial perturbation equations can be combined into a single hyperbolic, second-order partial differential equation (see Appendix~\ref{app:ax_eqs_time}). This equation is numerically solved using standard finite-difference schemes, such as the Lax–Wendroff method~\cite{Zenginoglu:2011zz,Duque:2023nrf,Duque:2023seg}. We write it schematically as
\begin{equation}
\left(-\hat{\mathcal W}+{\mathcal B}(r)\partial_r -\mathcal{V}(r)\right)\mathcal{Q} (t,r)=0\,,
\label{eq:Qcal}
\end{equation}
where
\begin{equation}
\mathcal{Q}(t,r)=\frac{q(r)}{r}h_1(t,r)\,,
\label{eq:Qcal_def}
\end{equation}
defines a convenient master function for time-domain evolution. The functions $q(r)$, ${\mathcal B}(r)$, and $\mathcal V(r)$ are provided in Appendix \ref{app:ax_eqs_time}, while 
\begin{align}
 \hat{\mathcal{W}}
 &=\frac{1}{4}\left(1-\frac{2m}{r}\right)\left[\left(1+\frac{2m}{r}\right)\partial_t^2\right .\nonumber\\
  &\left.-a\partial_r \left(\left(1-\frac{2m}{r}\right)a\partial_r\right)-\frac{4 m a}{r}\partial_r\partial_t+\frac{2ma}{r^2}\partial_t\right]\,, 
\end{align} 
is the wave operator in ingoing EF coordinates.
At large distances ($r \gg R_{\rm ga}$), $\mathcal Q(t, r)$ is approximately independent of $r$ and, for a certain time interval, can be represented as a superposition of damped sinusoids
\begin{equation}
   \mathcal Q(t,r)\approx\sum_{n} A_{n}\ e^{-\Im[\Upsilon_n]t}\cos(\Re[\Upsilon_n]t+\varphi_n)\,, \label{eq:ax:Qcal_TD_inf}
\end{equation}
where $A_n$ and $\varphi_n$ denote the amplitude and initial phase of the $n$-th mode, respectively. The QNM frequencies $\Upsilon_n$ are then obtained by fitting Eq.~\eqref{eq:ax:Qcal_TD_inf} to the numerical evolution of $\mathcal Q(t,r)$~\cite{Baibhav:2023clw}. This fit is repeated for several extraction radii ($r\gg R_{\rm ga}$) and time windows to assess the stability and consistency of the resulting frequencies.
\begin{figure*}
    \centering
    \includegraphics[width=2\columnwidth]{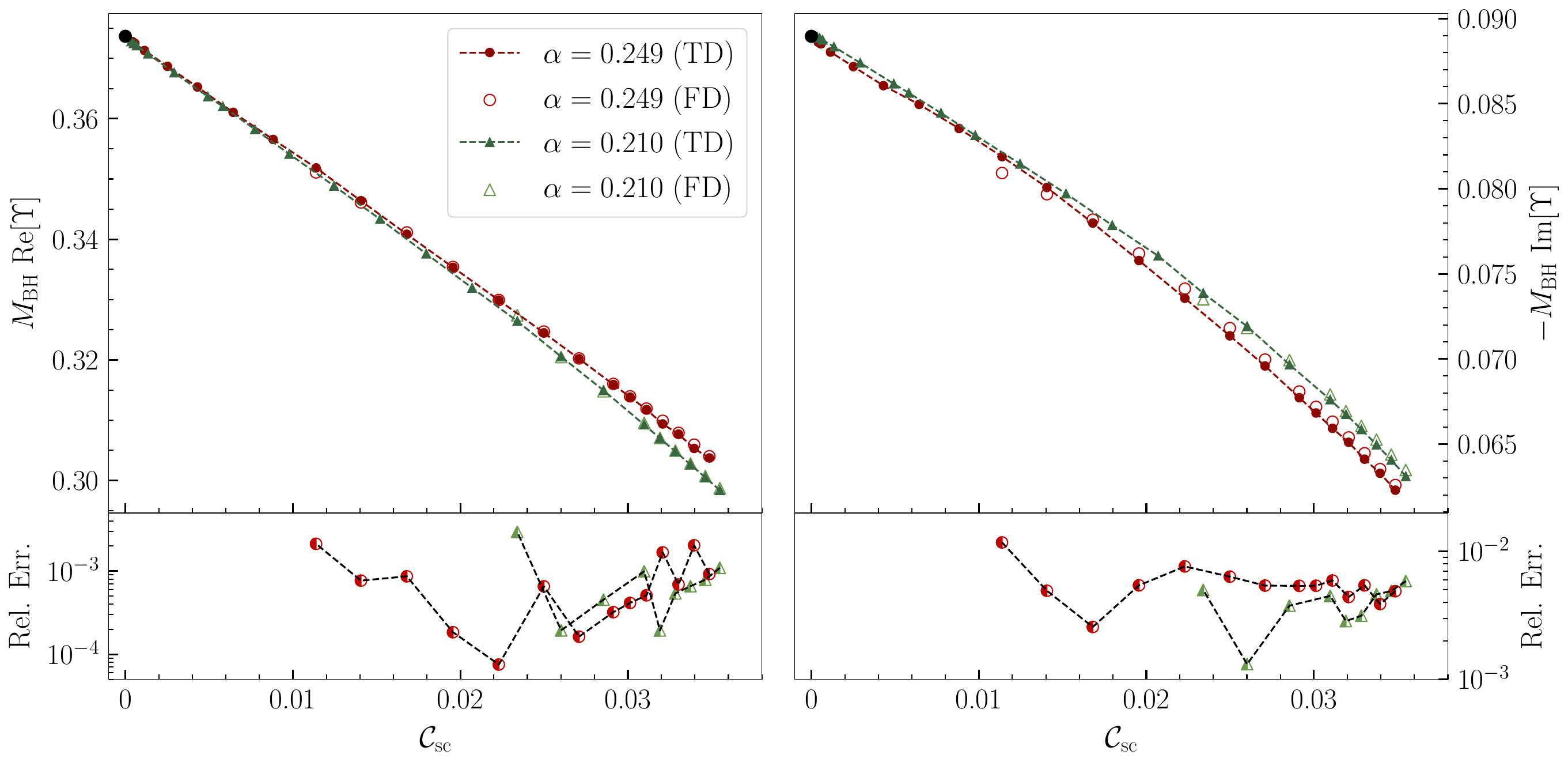}
    \caption{Comparison between time-domain (TD) and frequency-domain (FD) results. The top panels show the real and imaginary parts of the fundamental mode as functions of the compactness $\CSC$ for two values of $\alpha$. The Schwarzschild (vacuum) limit is marked by a black dot. The bottom panels display the relative differences between TD and FD computations (half-filled markers), which typically remain below $10^{-3}$ for the real part and $10^{-3}$ for the imaginary part.}
    \label{fig:validation}
\end{figure*}
\section{Results}\label{sec:results}
We now solve the perturbation equations in both the frequency~\eqref{eq:Q} and the time domain~\eqref{eq:Qcal}. Frequency-domain (FD) methods generally yield higher precision than time-domain (TD) approaches;~however, direct FD integration becomes numerically unstable at large radii~\cite{Chandrasekhar:1975zza}. This is particularly problematic for configurations like gravitational atoms, whose radial extent can reach tens to hundreds of Schwarzschild radii (scaling as $(\mu_{\rm s}\alpha)^{-1}$). In practice, we find that when $R_{\rm ga}\gtrsim 30\MBH$, FD integration becomes unreliable, and TD simulations are therefore required to explore the full parameter space.
To assess numerical accuracy, we performed convergence tests for both the time-domain (TD) and frequency-domain (FD) approaches and compared their outputs. For highly compact configurations, the fundamental QNM frequency $\Upsilon$ agrees between the two methods to within a relative error of $\sim 0.01\%$. At very low compactness values ($\mathcal{C}_{\rm sc}\lesssim 10^{-4}$), however, the relative frequency-shift away from the vacuum limit becomes comparable to the numerical uncertainty, preventing us from reliably determining a physically meaningful shift.
We therefore restrict our comparison to the regime where both methods remain accurate, namely $0.210 < \alpha < 0.249$ and $0.01 < \mathcal{C}_{\rm sc} < 0.04$. For smaller $\alpha$ or $\mathcal{C}_{\rm sc}$, the FD scheme loses accuracy, while for larger values the gravitational atom no longer supports radially stable configurations.
Figure~\ref{fig:validation} shows the real ({\it left panel}) and imaginary ({\it right panel}) parts of the fundamental mode $\Upsilon$ for $\alpha=0.249$ and $0.210$, together with TD results at lower compactness. The bottom panels display the relative difference between TD and FD results. The mean relative difference typically remains below $10^{-3}$ for the real part and $10^{-2}$ for the imaginary part, confirming the consistency and accuracy of both methods. As expected, the QNM frequency converges to the Schwarzschild (vacuum) limit, $\MBH\Upsilon_{\rm vac}\approx 0.3737-0.08896 i$~\cite{Leaver:1985ax}, when $\CSC \rightarrow 0$.
\begin{figure}[ht!]
     \centering     
     \includegraphics[width=0.98\columnwidth]{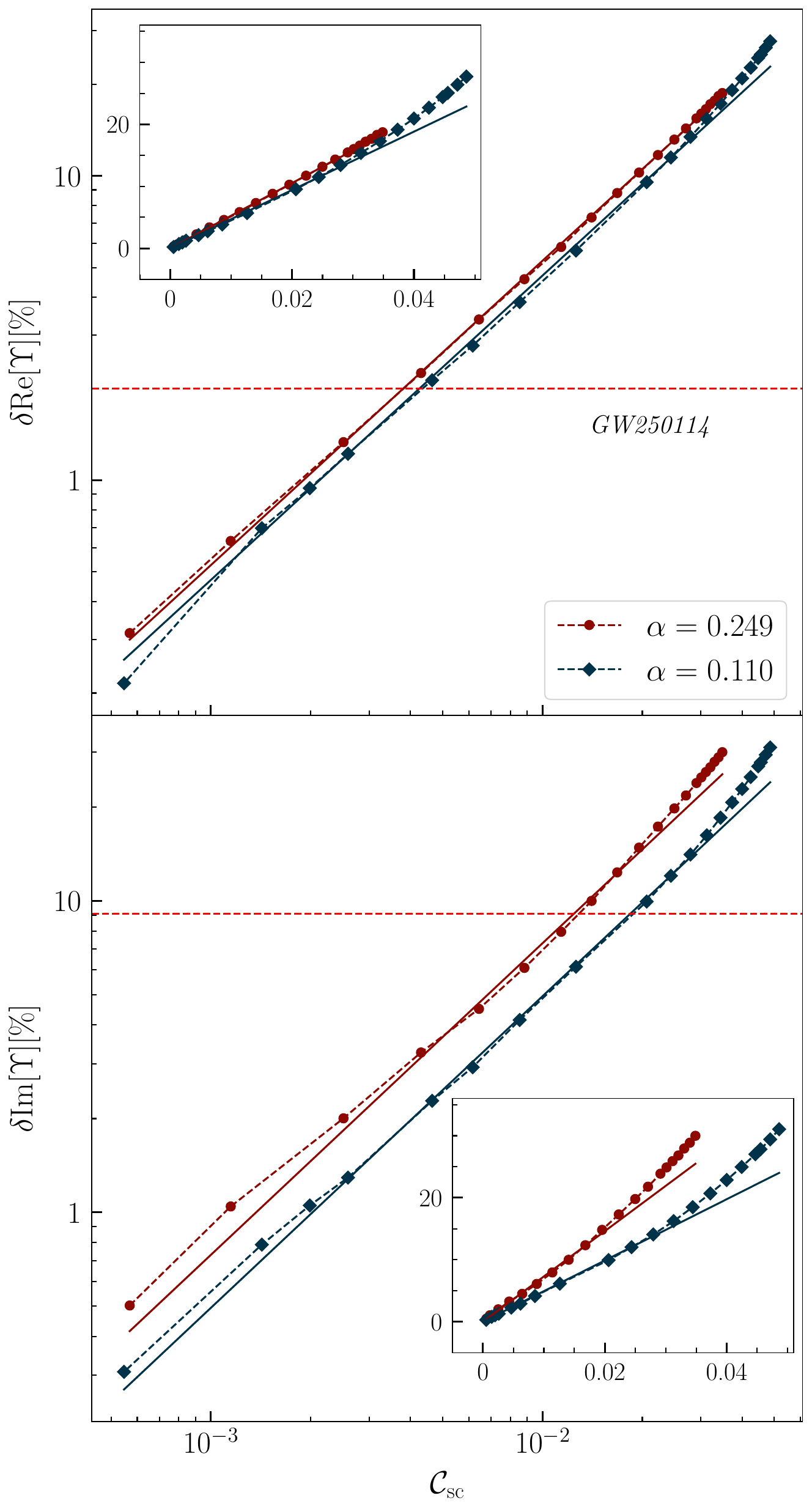}
     \caption{Percentage relative shift of the real ({\it top panel}) and imaginary ({\it bottom panel}) parts of the fundamental QNM as functions of the compactness for two values of $\alpha$:~$\alpha=0.249$ (red circles) and $\alpha=0.110$ (dark blue squares). For $\CSC\lesssim 3\times 10^{-2}$, the shift varies linearly with compactness;~the solid lines show the results of a linear fit. The insets show the same data on a linear scale, highlighting the deviation from the linear trend for $\CSC>10^{-2}$. The red dashed horizontal line marks the 90\% credible LVK constraint from the {\it GW250114} event~\cite{KAGRA:2025oiz}. Configurations above this line could, in principle, produce QNM shifts observable with the LVK detectors.}
     \label{fig:shift}
\end{figure}
\subsection{QNM shifts}\label{sec:fundamental_shift}
\begin{figure}[ht]
 \centering
 \includegraphics[width=\columnwidth]{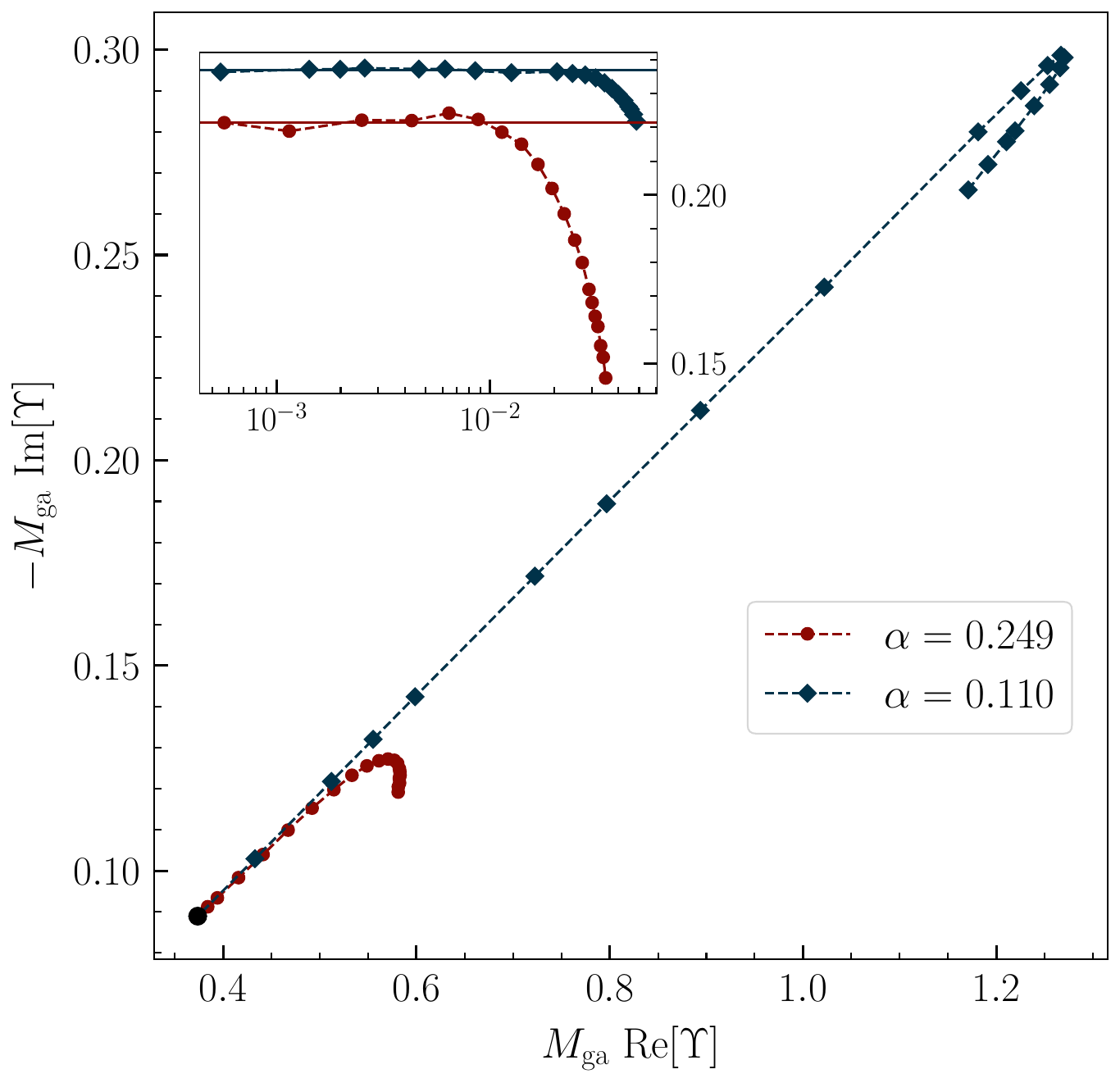}
 \caption{Imaginary versus real part of the fundamental mode for different scalar-cloud configurations. Results are shown for $\alpha=0.249$ (red circles) and $\alpha=0.110$ (dark blue squares). The black dot marks the vacuum limit, $\MBH\Upsilon_{\rm vac}$. The inset shows that for $\CSC<0.02$, the ratio $\mathscr{L}$~\eqref{eq:L_def} remains approximately constant.}
 \label{fig:ratio}
\end{figure}

To quantify the effect of the scalar cloud on the QNM spectrum, we compute the relative shifts in the real and imaginary parts of the fundamental mode as
\begin{equation}
       \delta\Re[\Upsilon]=1-\frac{\Re[\Upsilon]}{\Re[\Upsilon_{\rm vac}]}\,, \quad  \delta\Im[\Upsilon]=1-\frac{\Im[\Upsilon]}{\Im[\Upsilon_{\rm vac}]}\,.
\end{equation}
Figure~\ref{fig:shift} shows these shifts as functions of the compactness for two representative cases:~$\alpha=0.249$ (red circles) and $\alpha=0.110$ (dark blue squares), both computed from the TD analysis. As expected, both the real and imaginary components deviate increasingly from their vacuum values as the compactness grows. For moderately compact configurations ($\CSC \sim 0.01$), we find $\delta\Re[\Upsilon]\sim5\%$ and $\delta\Im[\Upsilon]\sim10\%$, reaching $\sim20\%$ and $\sim30\%$, respectively, for the most compact cases (see inset in Fig.~\ref{fig:shift}).
To provide a qualitative assessment of the relevance of these shifts, we compare them with current observational bounds on the fundamental mode obtained by the LVK collaboration from the recent ``loud'' event {\it GW250114}~\cite{KAGRA:2025oiz,LIGOScientific:2025obp}. In~\cite{LIGOScientific:2025obp}, deviations from the vacuum fundamental QNM frequency are parametrised as (cf.~Eq.~2 therein)
\begin{equation}
  f_{220}=f^{\rm vac}_{200}(1+\delta \hat f_{220})\,, \quad   \tau_{220}=\tau^{\rm vac}_{200}(1+\delta \hat \tau_{220})\,,
\end{equation}
where $f_{200}=\Re[\Upsilon]$ and $\tau_{200}=1/\Im[\Upsilon]$. Including pre-merger information and allowing for deviations in both the $220$ and $440$ modes, they report (cf. Fig.~4 in~\cite{LIGOScientific:2025obp}),
\begin{equation}
     \delta \hat f_{220}=0.02^{+0.02}_{-0.02}\,, \quad \delta\hat \tau_{200}=-0.01^{+0.10}_{-0.09}\,,\label{eq:90lvk}
\end{equation}
at the $90\%$ credible level.\footnote{We quote here only the results for the fundamental mode, which is the focus of this work. The correlation between the $220$ and $440$ modes is negligible~\cite{LIGOScientific:2025obp}.} Although these constraints were derived for a Kerr BH with spin $\chi\simeq0.7$, they provide a useful benchmark for estimating the detectability of environmental effects in the non-rotating case considered here.
In Fig.~\ref{fig:shift}, the red dotted lines indicate the $90\%$ credible LVK bounds from Eq.~\eqref{eq:90lvk}. Our results suggest that scalar clouds with compactness $\CSC\gtrsim3\times10^{-3}$ could, in principle, produce detectable shifts in the QNM frequency for events with a post-merger signal-to-noise ratio (SNR) comparable to that of {\it GW250114} (SNR $\simeq 40$). Detectability of both the real and imaginary shifts within the same credible interval would require $\CSC\gtrsim10^{-2}$. Since parameter uncertainties scale roughly as $1/\mathrm{SNR}$ (see, e.g.,~\cite{Berti:2005ys}), the minimum detectable compactness scales as
\begin{equation}
    \CSC(\rm SNR) \gtrsim \frac{10^{-2}\,\overline{\rm{SNR}}}{\rm SNR}\approx \frac{0.4}{\rm SNR}\,,
\end{equation}
where $\overline{\rm{SNR}}$ denotes the SNR of {\it GW250114}. 
As previously observed in studies of anisotropic fluid environments~\cite{Cardoso:2021wlq,Pezzella:2024tkf}, both $\delta\Re[\Upsilon]$ and $\delta\Im[\Upsilon]$ exhibit a nearly linear dependence on the compactness at small $\CSC$. The corresponding slopes, which show only mild dependence on $\alpha$, are summarised in Table~\ref{tab:shift} and are shown as solid lines in Fig.~\ref{fig:shift}.  For $\CSC \gtrsim 3\times10^{-2}$, deviations from the linear trend become evident (see inset of Fig.~\ref{fig:shift}).
\begin{table}[h!]
    \centering
    \begin{tabular}{c|c|c}
         $\alpha$&$\delta\Re[\Upsilon]$&$\delta\Im[\Upsilon]$  \\
         \hline
         0.249& $(5.26\pm0.01)\,\CSC$&$(7.30\pm0.09)\,\CSC$\\
         0.110&$(4.70\pm 0.03)\,\CSC$ & $(4.94\pm 0.03)\,\CSC$ 
    \end{tabular}
    \caption{Linear fits to the QNM frequency shifts as functions of the compactness for two values of $\alpha$. The linear approximation holds for $\CSC\lesssim3\times 10^{-2}$.}
    \label{tab:shift}
\end{table}
A useful reference point is provided by earlier work on anisotropic-fluid halos~\cite{Cardoso:2021wlq,Spieksma:2024voy,Pezzella:2024tkf}, where the compactness can be expressed in terms of the \emph{redshift} at the horizon $a^2_{\rm h}$. Empirically, we find that
\begin{equation}
    \CSC\approx \frac{1}{9.4}\left(1-a^2_{\rm h}\right)\,,
\end{equation}
which, to a good approximation, is independent of $\alpha$. At small compactness and for fixed $a_{\rm h}$, the frequency shifts induced by the scalar cloud is thus comparable to those from the halo~\cite{Pezzella:2024tkf}.
Finally, Fig.~\ref{fig:ratio} shows the imaginary part of $\Upsilon$ as a function of its real part, both normalised by the total mass of the gravitational atom, $M_{\rm ga}$, rather than by the BH mass $\MBH$. This normalisation removes any residual dependence on the mass of the gravitational atom and thus isolates the effect of the compactness.
For $\CSC\lesssim 0.02$, the ratio
\begin{equation}
 \mathscr{L}\equiv -\frac{\Im[M_{\rm ga}\Upsilon]-\Im[\MBH\Upsilon_{\rm vac}]}{\Re[M_{\rm ga}\Upsilon]-\Re[\MBH \Upsilon_{\rm vac}]}\,,
\label{eq:L_def}
\end{equation}
is nearly constant and independent of the compactness, making it a convenient diagnostic of environmental effects. Its dependence on $\alpha$ encodes the imprint of the scalar field mass $\hbar \mu_{\rm s}$ (see inset in Fig.~\ref{fig:ratio}). Specifically, we find $\mathscr{L}\approx 0.2209\,\pm\,7\times 10^{-4}$ for $\alpha=0.249$ and $\mathscr{L}\approx 0.2369\,\pm\,10^{-4}$ for $\alpha=0.110$. If the remnant BH mass can be independently measured -- for example, through inspiral-only analyses~\cite{Ghosh:2017gfp} or electromagnetic observations~\cite{Remillard:2006fc} -- then the observed QNM frequency shift of a gravitational atom could provide a direct probe of the scalar field mass.
\subsection{Light-ring frequency, Lyapunov exponent and surface gravity}
\begin{figure}
    \centering
    \includegraphics[width=0.98\linewidth]{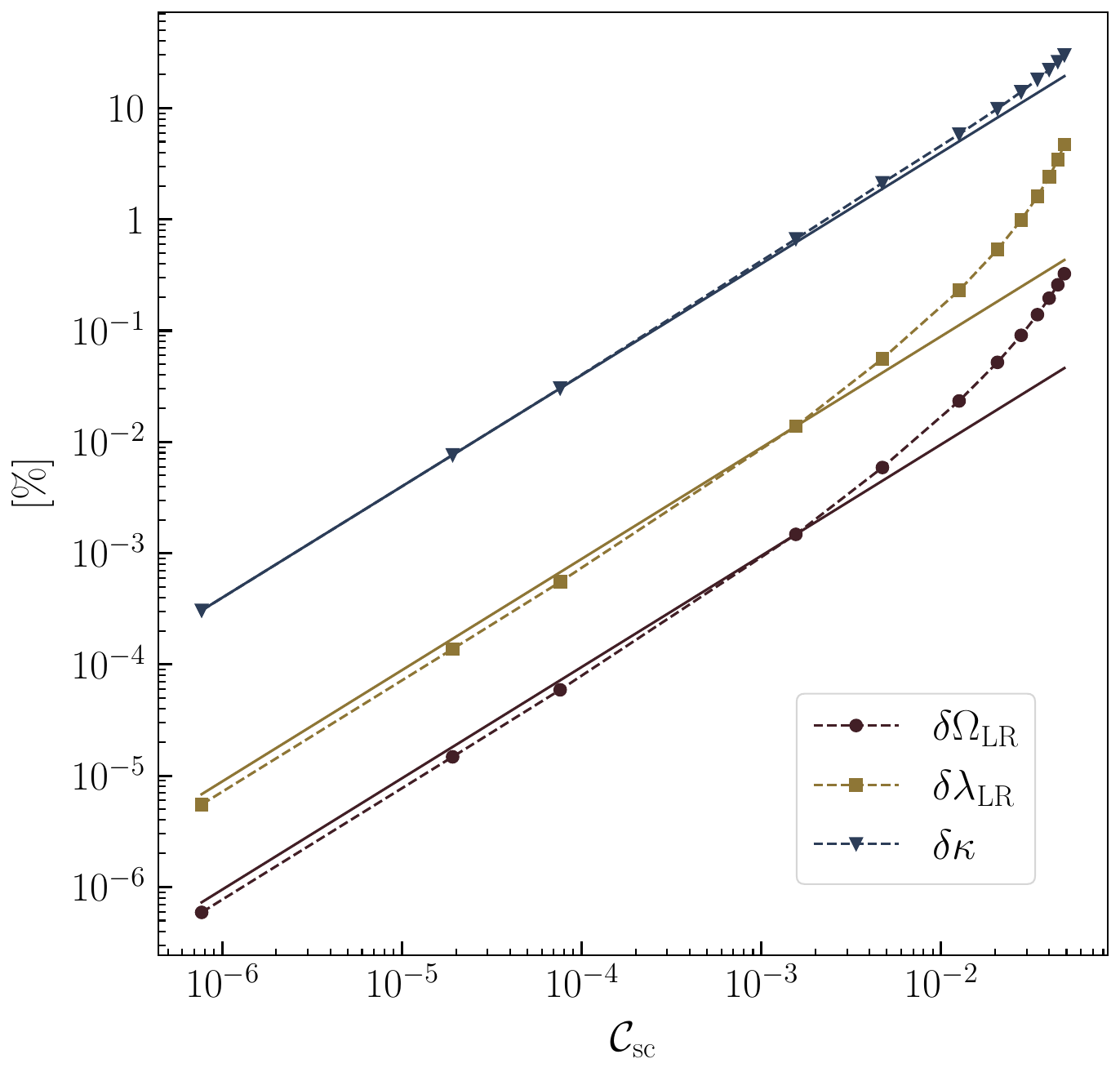}
    \caption{Relative percentage shifts of the light-ring frequency $\Omega_{\rm LR}$, Lyapunov exponent $\lambda_{\rm LR}$ and surface gravity $\kappa$ induced by the presence of a scalar cloud, shown relative to the vacuum case for $\alpha=0.110$. Solid lines denote linear fits valid for $\CSC\lesssim 10^{-4}$.}
    \label{fig:lightring}
\end{figure}

To gain further physical insight into the system, we compute three background quantities closely related to the QNM spectrum~\cite{Berti:2009kk}:~the light-ring frequency, the Lyapunov exponent, and the surface gravity.
In vacuum, the so-called {\it eikonal limit} ($\ell\gg1$) establishes a direct connection between the QNM spectrum and the properties of the photon sphere. In this regime, the QNM frequencies of Kerr BHs can be approximated by~\cite{Cardoso:2008bp}
\begin{equation}
    \Upsilon\approx \Omega_{\rm LR}\ell -i \left(n+\frac{1}{2}\right)|\lambda_{\rm LR}|\,,\label{eq:eikonal_limit}
\end{equation}
where $n$ is the overtone number, $\Omega_{\rm LR}$ the orbital frequency of the light ring, and $\lambda_{\rm LR}$ the Lyapunov exponent characterising the instability timescale of the photon orbit. Following~\cite{Cardoso:2008bp}, the light-ring radius $r=r_{\rm LR}$ satisfies\footnote{In the quasi-stationary approximation, the metric depends only on $r$, which coincides with the Schwarzschild radial coordinate. We therefore use the same expressions as in~\cite{Cardoso:2008bp}.} 
\begin{equation}
\frac{a'}{a}\biggl|_{r = r_{\rm LR}}=\frac{rm'+r-3m}{r(r-2m)}\biggl|_{r = r_{\rm LR}}\,,
\end{equation}
which reduces to $r_{\rm LR}=3\MBH$ in the vacuum limit. The corresponding light-ring frequency and Lyapunov exponent are
\begin{equation}
\begin{aligned}
   \MBH\Omega_{\rm LR}&=\frac{a(r_{\rm LR})}{r_{\rm LR}}\sqrt{1-\frac{2 m(r_{\rm LR})}{r_{\rm LR}}}\,,\\
    \MBH\lambda_{\rm LR}&=\frac{a}{r^2}\biggl[r^2 \left(3 m'^2+r m''\right)-2 r m \left(3 m'+r m''\right)\\
&+3 m^2-r^2 (r-2 m)^2 \frac{a''} {a}\biggl]^{1/2} \ \biggl|_{r = r_{\rm LR}}\,,
\end{aligned}
\end{equation}
both of which reduce to $1/(3\sqrt{3}\MBH)$ in Schwarzschild. 
For high overtones ($n\gg 1$) instead, the QNM spectrum is well approximated by~\cite{Padmanabhan:2003fx,Motl:2003cd,Motl:2002hd,Medved:2003rga,Nollert:1993zz,Andersson:1993ak}
\begin{equation}
    \Upsilon\approx \kappa \left[i \left(n+\frac{1}{2}\right)+\frac{\ln(3)}{2}\right]\,,
    \label{eq:overtone_limit}
\end{equation}
where $\kappa$ is the surface gravity of the BH, defined as~\cite{Wald:1984rg}
\begin{equation}
    \kappa^2=-\frac{1}{2}\nabla^\mu t^\nu \nabla_{\mu}t_{\nu}\,,
\end{equation}
with $t^\nu$ the timelike Killing vector. For gravitational atoms, this becomes
\begin{equation}
    \kappa=\frac{a_{\rm h}}{4\MBH}\left[1-4\mathcal{A}^2\alpha^2\left(1+2\frac{|\Omega|^2}{\mu_{\rm s}^2a^2_{\rm h}}\right)\right]\,,
\end{equation}
which reduces to the Schwarzschild value $\kappa_{\rm vac}=(4\MBH)^{-1}$ in vacuum.
Figure~\ref{fig:lightring} shows the relative shifts of $\Omega_{\rm LR}$, $\lambda_{\rm LR}$ and $\kappa$ as functions of the compactness for $\alpha=0.110$.
Similar to before, in the small-compactness regime ($\CSC\lesssim10^{-4}$), they are well described by linear fits:
\begin{equation}
\begin{aligned}
    \delta\Omega_{\rm LR}&=1-\frac{\Omega_{\rm LR}}{\Omega^{\rm vac}_{\rm LR}}\simeq 0.0095\,\CSC\,, \\
    \delta \lambda_{\rm LR}&=1-\frac{\lambda_{\rm LR}}{\lambda^{\rm vac}_{\rm LR}}\simeq  0.088\,\CSC\,, \\
    \delta\kappa&=1-\frac{\kappa}{\kappa_{\rm vac}}\simeq 3.99\,\CSC\,.   
\end{aligned}
\label{eq:LR_shift}
\end{equation}
Comparing Eq.~\eqref{eq:LR_shift} with the QNM shifts reported in Table~\ref{tab:shift} shows that the presence of the scalar cloud completely spoils the eikonal correspondence~\eqref{eq:eikonal_limit}:~the QNM frequency shifts are up to two orders of magnitude larger than the light-ring predictions.\footnote{For $\alpha=0.249$, we find $\delta\Omega_{\rm LR}\simeq 0.45,\CSC$, $\delta\lambda_{\rm LR}\simeq 1.93,\CSC$, and $\delta\kappa\simeq 6.49,\CSC$.} This breakdown of the geodesic (eikonal) correspondence is not unexpected. As noted in~\cite{Alcubierre:2024mtq}, self-gravitating gravitational atoms interpolate between BHs and boson stars.  The geodesic correspondence is known to fail for boson stars, which lack a light ring unless they are \emph{ultracompact}~\cite{Cunha:2015yba}. In the limit of small $\alpha$, gravitational atoms increasingly resemble boson stars, leading to larger deviations from Eq.~\eqref{eq:eikonal_limit}.
Conversely, the surface-gravity shifts are still comparable to the QNM frequency shifts across the entire parameter space, including for the most compact configurations where the frequency–compactness relation is no longer linear. This suggests that QNMs are more sensitive to the near-horizon geometry than to the photon-sphere structure. To further illustrate this, we compare our results again with those for anisotropic-fluid halos. A straightforward calculation shows that the surface gravity $\kappa_{\rm H}$ for a halo with stress-energy tensor $T_{\mu}^\nu=\rm{diag}\{-\rho,p_{\rm r},p_{\rm t},p_{\rm t}\}$ is 
\begin{equation}
\kappa_{\rm H}=\frac{a_{\rm h}}{2 r_{\rm h}} \left(8 \pi  r_{\rm h}^2\ p_{\rm r, h}+1\right) \,,\label{eq:surface_gravity_shift}
\end{equation}
where $p_{\rm r,h}$ is the radial pressure evaluated at the horizon. Since $p_{\rm r,h}=0$ for anisotropic fluids, Eq.~\eqref{eq:surface_gravity_shift} implies that the surface gravity scales directly with $a_{\rm h}$ and thus as the shift of the fundamental mode~\cite{Pezzella:2024tkf}. The surface gravity correspondence therefore does hold for both matter halos and scalar cloud environments, reinforcing analytical arguments suggesting that Eq.~\eqref{eq:overtone_limit} still holds in beyond vacuum BHs~\cite{Medved:2003rga}.
\section{Conclusions}\label{sec:conclusions}
We have extended the BH spectroscopy programme to {\it gravitational atoms}:~BHs surrounded by long-lived, self-gravitating scalar clouds. Through a fully relativistic analysis of the axial sector, we computed the fundamental QNM in both the frequency and time domains, finding consistent results between the two. As expected, the presence of the scalar cloud induces shifts in both the real and imaginary parts of the fundamental QNM frequency. Although the ringdown signal is expected to be largely unaffected by most astrophysical environments~\cite{Spieksma:2024voy}, our results show that highly compact configurations can leave detectable imprints on the QNM spectrum. For instance, for signal-to-noise ratios comparable to that of {\it GW250114}, scalar clouds with $\CSC \gtrsim3\times10^{-3}$ could produce deviations from the vacuum spectrum detectable with the current LVK sensitivity. 
Such compactness values could possibly arise through superradiant instabilities~\cite{Brito:2015oca} or through relaxation and gravitational cooling processes~\cite{Hui:2021tkt}. The main caveat, however, concerns the survival of the cloud throughout the coalescence. As the inspiral proceeds, the cloud is expected to experience depletion due to the binary's motion. Recent numerical relativity simulations of BH binaries in scalar-field environments suggest that, owing to their wave-like nature, such clouds may be depleted less efficiently during the merger than particle-like distributions~\cite{PhysRevD.103.044032,Aurrekoetxea:2024cqd,Cheng:2025wac}, and may even induce detectable phase shifts in the late inspiral. Nonetheless, longer simulations, together with a better understanding of the cloud’s past evolution~\cite{Tomaselli:2024bdd,Tomaselli:2024dbw}, are necessary to draw firmer conclusions.
We also examined background quantities related to the QNM spectrum, including the light-ring frequency, Lyapunov exponent, and surface gravity. While the eikonal (geodesic) correspondence between the light ring and QNM frequencies breaks down in the presence of the scalar cloud, the surface-gravity estimate seems robust, suggesting that QNMs are mostly sensitive to the near-horizon geometry rather than to the structure of the photon sphere.
Our findings motivate several directions for future investigation. Extending the analysis to the polar sector is particularly important, as scalar and gravitational perturbations couple there, producing an additional family of modes imprinted in the GW signal. Moreover, given recent suggestions that LVK events may already be affected by scalar-field environments during the inspiral stage~\cite{Roy:2025qaa}, one should also study in detail the post-merger phase, determining how much of the cloud survives, the corresponding shifts to the QNM spectrum, and whether these signatures are consistent with current observations.  

\noindent {\bf Acknowledgments.} 
MDR thanks G.~Antoniou, A.~Giorgieri, and C.F.B.~Macedo for useful discussions. T.S. acknowledges support from a Royal Society University Research Fellowship (URF-R1-231065).
The Center of Gravity is a Center of Excellence funded by the Danish National Research Foundation under grant No. 184.
We acknowledge support by VILLUM Foundation (grant no. VIL37766).
V.C.\ is a Villum Investigator and a DNRF Chair.  
V.C. acknowledges financial support provided under the European Union’s H2020 ERC Advanced Grant “Black holes: gravitational engines of discovery” grant agreement no. Gravitas–101052587. 
Views and opinions expressed are however those of the author only and do not necessarily reflect those of the European Union or the European Research Council. Neither the European Union nor the granting authority can be held responsible for them.
This project has received funding from the European Union's Horizon 2020 research and innovation programme under the Marie Sklodowska-Curie grant agreements No 101007855 and No 101131233.
This work is supported by Simons Foundation International \cite{sfi} and the Simons Foundation \cite{sf} through Simons Foundation grant SFI-MPS-BH-00012593-11.
\appendix
\section{Quasi-stationary self-gravitating gravitational atoms}\label{app:bkg}
In this appendix, we provide the details of the construction of the \emph{self-gravitating gravitational atoms}: quasi-stationary, spherically symmetric background solutions used throughout the main text.
Substituting Eqs.~\eqref{eq:Phi0}-\eqref{eq:bkg_metric} into the Einstein–Klein–Gordon equations yields the coupled system for the metric functions~\eqref{eq:bkgEQSCompactForm} and the scalar field:
\begin{equation}
    f \psi ''=\mathcal{C} \psi +\mathcal{B}\psi'\,,\quad \text{where}\quad f=1-\frac{2 m}{r}\,, \label{eq:psi}
\end{equation}
with the coefficients,
\begin{subequations}\label{Eq:BC}
\begin{eqnarray}
\mathcal{B} &=& -\frac{2}{r}\left( 1 - \frac{m}{r} \right) - \frac{4m}{r}\frac{i\Omega}{a}
\nonumber\\
 &+& \frac{e^{2\sigma t} r}{\gamma^2}\biggl\{ 
  (\gamma^4-1)\left| \frac{\Omega}{a}\psi \right|^2 + \gamma^2 \mu_{\rm s}^2 |\psi|^2\nonumber\\
   &+& (1+\gamma^2)f\,\Re\!\left[ \frac{-i\Omega^*}{a}\psi^*\psi' \right]
  \biggr\}\,,
\label{Eq:B}\\
\mathcal{C} &=& \mu_{\rm s}^2 + \gamma^2\left( \frac{i\Omega}{a} \right)^2 - \frac{2m}{r^2}\frac{i\Omega}{a}
\nonumber\\
 &-&   e^{2\sigma t} r\biggl\{ \left| \frac{i\Omega}{a}\psi \right|^2 + \mu_{\rm s}^2 |\psi|^2
   + f\left| \psi' \right|^2\nonumber\\
  & -& f\,\Re\!\left[ \frac{-i\Omega^*}{a}\psi^*\psi' \right]
\biggr\} \frac{i\Omega}{a}\,,
\label{Eq:C}
\end{eqnarray}
\end{subequations}
where $\gamma=\sqrt{1+2m/r}$. The functions $F$ and $G$ in Eq.~\eqref{eq:bkgEQSCompactForm} read
\begin{align}
 F(r)&= a r  \left| \frac{i\Omega \psi}{a}-\psi'\right| ^2\,,\label{eq:a} \\
G(r) &=\frac{   r^2}{2}\left[ | \psi| ^2 \left(\mu_{\rm s} ^2  +\frac{|\Omega| ^2}{a^2} \gamma^2\right)+ f\left| \psi '\right| ^2  \right]\,. \label{eq:m}
\end{align}
In addition, Einstein’s equations imply
\begin{equation}
\partial_t m=r^2 e^{2 \sigma  t} \left( \frac{2 m|\Omega \psi |
   ^2}{ar}-f\,\Im\!\left[\Omega \psi  {\psi
   '}^*\right]\right) \ .
   \label{eq:dt_m}
\end{equation}
Quasi-stationary solutions exist for times $t\lesssim 1/(2\sigma)$~\cite{Alcubierre:2024mtq}. However, even when $2\sigma t\to 0$, the right-hand side of~\eqref{eq:dt_m} does not vanish identically due to our choice of coordinates. In the limit $\sigma\to0$, the no-hair theorems forbid self-gravitating gravitational atoms:~the solution must correspond to either a vacuum BH or a boson star~\cite{Pena:1997cy}. In this limit, the mass must be time-independent. Indeed, transforming Eq.~\eqref{eq:dt_m} to Schwarzschild coordinates (cf.~Eq.~36 in~\cite{Barranco:2017aes}) yields $\partial_t m = 0$ for $\sigma\to0$ when $\psi\in\mathbb{R}$, thus satisfying the no-hair theorem~\cite{Pena:1997cy}. For all configurations studied, we verify {\it a posteriori} that $\partial_t m/(2m\sigma)\lesssim1$.
To then construct the background solution, we solve the coupled ODEs for the metric functions~\eqref{eq:bkgEQSCompactForm} and the scalar field~\eqref{eq:psi} numerically with suitable boundary conditions. At the horizon $r = r_{\rm h}$, all fields are expanded as power series,\footnote{Alcubierre et \textit{al.} \cite{Alcubierre:2024mtq} made these expansions defining a new variable $z$ which makes the Klein-Gordon equation well-defined at $2m=r$. However, once we know that a solution exists, we can do these expansions in terms of $r$ instead of $z$.}
\begin{equation}
\begin{aligned}
    m(r)&=\frac{r_{\rm h}}{2}+\sum_{n=1}^{N}m_n(r-r_{\rm h})^n\,,\\
    a(r)&=1+\sum_{n=1}^{N}a_n(r-r_{\rm h})^n\,,\\
    \psi(r)&=\mathcal{A}+\sum_{n=1}^{N}\psi_n(r-r_{\rm h})^n\,. 
\end{aligned}
\label{eq:bkg_exp_hor}
\end{equation}
We tested different truncation orders $N$ and found that in practice $N=0$ is sufficient to numerically compute the background. However, to compute the horizon series coefficients needed later when imposing the boundary conditions for the perturbations (the coefficients $q_n^{\rm h}$ in Eq.~\eqref{eq:ax:bounds}), we use $N=3$ so that the near-horizon expansions match the Schwarzschild limit smoothly. At spatial infinity ($r\to\infty$) the scalar field must decay exponentially as 
\begin{equation}
   \psi(r) \to \psi_\infty(r)\equiv e^{-k_{\rm s} r} r^{-1-M \left(2 i k_{\rm s} \overline{\Omega}-2 \overline{\Omega }^{\,2}+\mu_{\rm s}^2\right)/k_{\rm s}}\,, 
\label{eq:psi_infty}
\end{equation}
where $k_{\rm s}=\sqrt{\mu_{\rm s} ^2-\overline{\Omega}^{\, 2}}$, $a_\infty\equiv a(r\to\infty)$ and $\overline{\Omega}\equiv\Omega/a_\infty$ is the frequency measured by a static observer at infinity.
We integrate the ODEs from $r_1=r_{\rm h}+\epsilon$ (with $\epsilon\sim10^{-4}$ to avoid numerical issues at the horizon) to a large radius $r_\infty\gtrsim 50M_{\rm BH}\gg r_{\rm H}$. For fixed $(\mathcal{A},\alpha)$ we determine the complex eigenfrequency $\Omega$ by a shooting method such that
\begin{equation}
    \left(\psi'\psi_\infty-\psi\psi'_\infty\right)\!\bigl|_{r = r_\infty}=0\,,
\end{equation}
and $|\psi(r)|\neq 0$ for $r\ge r_{\rm h}$. After finding the correct $\Omega$ we rescale $a(r)$ and $\Omega$ so that $a(r\to\infty)=1$ and $\Omega=\bar\Omega$.
Figure~\ref{fig:bkg1} (main text) shows the radial profiles $m(r)$, $a(r)$ and the real and imaginary parts of $\psi(r)$ for a representative configuration ($\alpha=0.249$, $\mathcal{A}=0.401$). Table~\ref{tab:bkg_results} lists additional background parameters (frequency $\Omega$, characteristic radius $R_{\rm ga}$, ADM mass $M_{\rm ga}$ and compactness $\mathcal{C}_{\rm sc}$) for a few choices of $(\mathcal{A},\alpha)$.
\begin{figure}
    \centering
    \includegraphics[width=\linewidth]{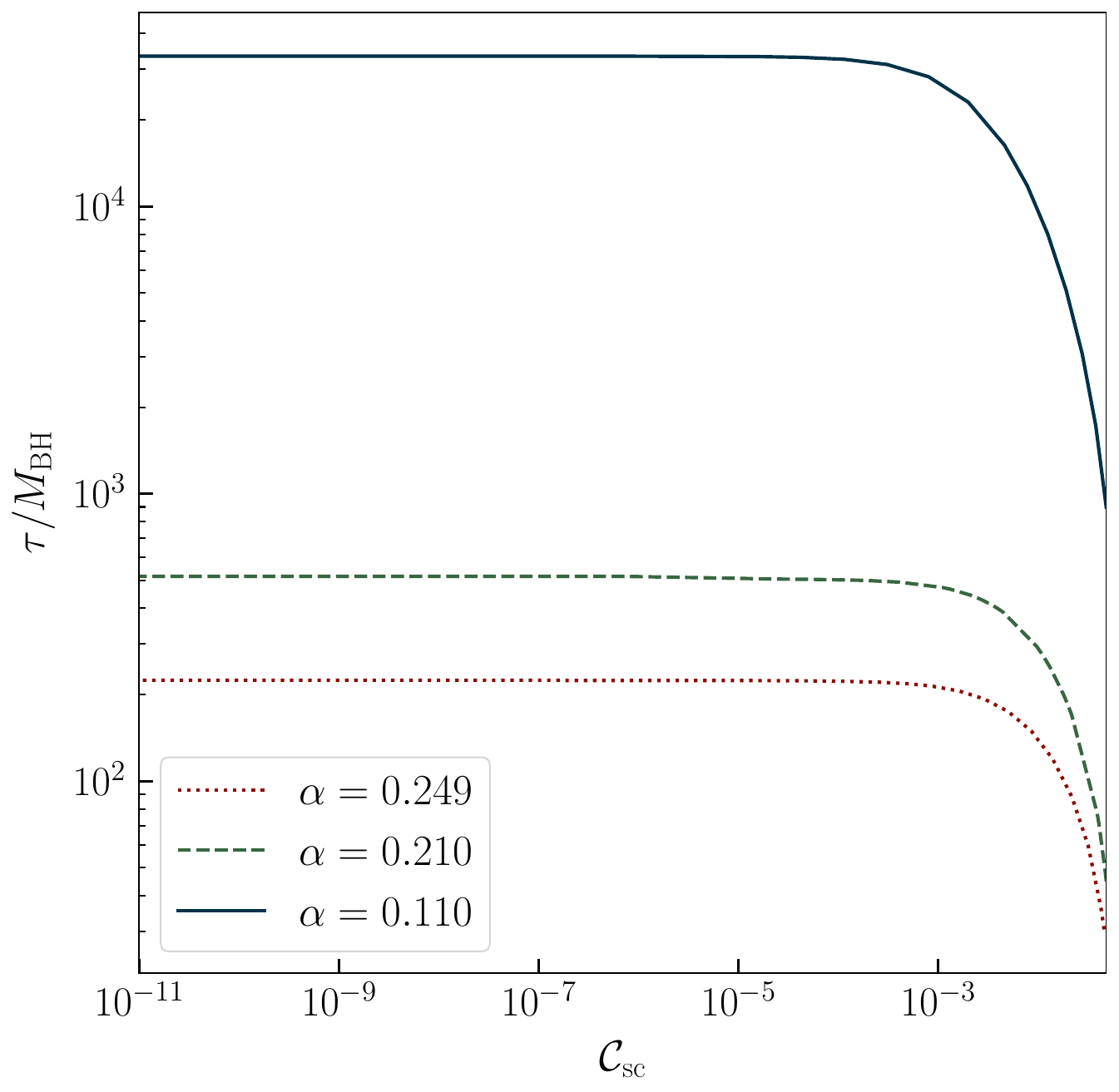}
    \caption{Decay timescale of the gravitational atom as a function of the compactness for three different values of $\alpha$. In the vacuum limit, the frequency coincides with the fundamental quasi-bound state of non-rotating BHs.}
    \label{fig:timedecay}
\end{figure}
In Fig.~\ref{fig:timedecay} we show the characteristic damping time of the cloud, $\tau=1/\sigma$, as a function of $\CSC$ for $\alpha=\{0.11,0.21, 0.249\}$. At $\CSC<10^{-3}$, the backreaction from the cloud is negligible and the scalar field profile is approximately that of a quasi-bound state configuration in vacuum (with a frequency that coincides with the fundamental mode~\cite{Detweiler:1980uk}). As can be seen in the figure, the time decay in this regime becomes approximately independent of the compactness. For $\CSC>10^{-3}$ instead, more compactness configurations have a shorter lifetime, though always remaining much larger than the dynamical timescale $\sim\MBH$ for all values of the compactness explored in this work.
\begin{table*}[t]
    \begin{tabular}{cc|cccc}
         $\alpha$&$\mathcal{A}$&   $\Omega \MBH$  & $R_{\rm ga}/\MBH$ & $M_{\rm ga}/\MBH$ & $C_{\rm sc}$\\
        \hline
       \multirow{3}{*}{$0.21$} & $5.013\times 10^{-7}$&$0.9731+0.009224 i$ & 2 & 1 &$\lesssim 10^{-11}$\\
       &  $5.329\times 10^{-2}$ & $0.9686 +0.01036   i$ &  44.32 & 1.110  &$1.5\times 10^{-3}$\\
         &  $3.528\times 10^{-1}$ & $0.8861 +0.04182  i$ & 32.00& 2.115  &$3.115\times 10^{-2}$\\
         \hline
        \multirow{3}{*}{$0.249$} & $5.013\times 10^{-7}$&$0.9649+0.01789 i$ & 2 & 1 &$\lesssim 10^{-12}$\\
       &  $2.395\times 10^{-2}$ & $0.9640 +0.01818  i$ &  10.90 & 1.015  &$2.7\times 10^{-4}$\\
         &  $2.013\times 10^{-1}$ & $0.9248 +0.03366  i$ & 31.76& 1.542  &$1.415\times 10^{-2}$
    \end{tabular}
        \caption{Parameters of the background solutions for different values of $\mathcal{A}$ and $\alpha$.}
    \label{tab:bkg_results}
\end{table*}
\section{Master equations}\label{app:ax_eqs}
In this appendix, we collect the perturbation equations used in Section~\ref{sec:perturbations} and outline how the metric and scalar field perturbation equations reduce to single second-order master equations in both the frequency~\eqref{eq:Q} and time domain~\eqref{eq:Qcal}.
In the Regge–Wheeler gauge~\cite{Regge:1957td} axial metric perturbations are expanded in axial vector harmonics $S_A$ and depend on two functions $\bar h_0(t,r)$ and $\bar h_1(t,r)$:
\begin{equation}
    h^{\rm ax}_{jA}= \bar{h}_j S_A\,, \quad j=\{0,1\}\,, \quad \mbox{ and } \quad A=\{\theta,\varphi\}\,,
\end{equation}
It is convenient to redefine these fields so that the vacuum Regge–Wheeler limit is manifest and a single second-order equation arises even for non-vacuum backgrounds. We therefore set
\begin{equation} 
\begin{aligned}
    h_0(t,r)&=\bar{h}_0(t,r)\,,\\
     h_1(t,r)&=\bar{h}_1(t,r)+\frac{2 m(r)}{a(r) (r-2 m(r))}\bar{h}_0(t,r)\,.
\end{aligned}
\label{eq:ax_redef}
\end{equation}
With this redefinition the linearised Einstein equations for the axial sector reduce to a coupled system for $(h_0,h_1)$ which we now present in the frequency and time domain.
\subsection{Frequency domain} \label{app:ax_eqs_frequency}
Using the frequency ansatz [cf.~Eq.~\eqref{eq:ax:f_fd}],
\begin{equation}
    h_{j}(t,r)=\sum_{n} \tilde{h}_j(r,\Upsilon_n)\,e^{-i\Upsilon_n t}\,,\quad j=\{0,1\}\,,
\end{equation}
and dropping the harmonic indices $(\ell,m)$ as well as the dependencies on $(t,r)$ for compactness, the linearised Einstein equations become
\begin{equation}
\begin{aligned}
&\tilde h_1\left\{    f|\psi ' |^2+   \mathcal{M}_{\rm s}^2 |\psi|^2+\frac{\ell ^2+\ell -2}{2 r^2 }+\frac{m\zeta}{ r a }-\frac{
   \Upsilon ^2}{2  a^2 f}\right.\\
   &\left.-\frac{3 a' m'}{r a}+\frac{(m+r) a'}{r^2 a}+\frac{f a''}{ a}-\frac{m''}{r}\right\}\\
   &+\frac{i  \Upsilon  \tilde h_0'}{2 a^2 f}+\frac{\Upsilon  \tilde h_0  ( \Upsilon  m-i a f)}{a^3rf^2}=0\,,\\
   &\left(\frac{f a'}{a}-\frac{2 i \Upsilon  m}{r a}-\frac{2 m'}{r}+\frac{2 m}{r^2}\right)\tilde h_1+\frac{i  \Upsilon  \tilde h_0}{a^2 f}+f \tilde h_1'=0\,.
\end{aligned}
\label{eq:h01}
\end{equation}
Here we introduced
\begin{equation}
    \mathcal{M}_{\rm s}= \mu_{\rm s}   ^2 -\frac{|\Omega|^2 }{a^2}\gamma^2\,,\quad \text{and}\quad \zeta=-4\,\Im[\Omega  \psi  {\psi'} ^*]\,.
\end{equation}
Following the vacuum case~\cite{Regge:1957td}, Eqs.~\eqref{eq:h01} can be combined into a single second-order master equation for the variable
\begin{equation}
Q(r)=\frac{p(r)\tilde h_1(r)}{r},
\end{equation}
where $p(r)$ reads
\begin{equation}
\begin{aligned}
&p(r)=e^{i\Upsilon(\hat r-r_*)}\\
&\times \exp \left(\int \left[r \left(r a'-2 a m'\right)+2 m \left(-r a'+a\right)\right]\frac{1}{r^2 a f} \, \dd r\right)\,.
\end{aligned}
\label{eq:p}
\end{equation}
With this definition, $Q(r)$ satisfies a Schrödinger-like equation (Eq.~\eqref{eq:Q} in the main text), with a potential given by
\begin{align}
    V(r)=&\frac{2 a f \left(a-3 r a'\right) m'}{r^2}+2   f |\psi|^2 a^2\mathcal{M}_{\rm s}\nonumber\\
    &+\frac{af}{r^3} \left[r (4 m+r) a'+a (r \ell  (\ell +1)-6 m)\right]\nonumber\\
    &+2 a f^2 a''-\frac{2 a^2f m''}{r}+2    a^2 f^2 |\psi '|^2+\frac{   \zeta a f m }{r}\,.
    \label{eq:V}
\end{align}
\subsection{Time domain}\label{app:ax_eqs_time}
In the time domain, the Einstein's equations read
\begin{equation}
\begin{aligned}
&h_1\left(f\frac{a'}{a}+\frac{2m}{r^2}-\frac{ m'}{r}\right)-\frac{\dot h_0}{a^2f}+\frac{2 m
   \dot h_1}{r a}+f h'_1=0 \,,\\
  & h_0\left\{\frac{ a''}{a}-\frac{3 a'm'}{r a}+\frac{m \left(3 a'-2a''r+r\zeta \right)}{r^2 a}\right. \\
   &\left. +rf|\psi '|^2-\frac{m''}{r}+\frac{\ell ^2+\ell}{2 r^2}+ \mathcal{M}_{\rm s}
   |\psi|^2 -\frac{2m}{r^2}
   \right\} \\
   &+\dot h_0 \left(\frac{2 m a'}{r a^2}+\frac{m-r m'}{r^2f a}\right)-f\left(\frac{a'}{2a}-\frac{1}{r}\right) \dot h_1 \\
   &+\frac{fa'h'_0}{2  a}-\frac{2 m^2 \ddot h_0}{r^2 a^2
   f}-\frac{2 m \dot h'_0}{r a}+\frac{m \ddot h_1}{r a}-\frac{f}{2} 
   h''_0+\frac{f}{2} \dot h'_1=0 \,,\\
   &h_1 \left\{\frac{a' \left(-3 r m'+m+r\right)}{r^2 a}+\frac{r a''-2m a''+ m \zeta}{r a}\right.\\
   &\left.+f |\psi '|^2-\frac{m''}{r}+\mathcal{M}_{\rm s}|\psi|^2+\frac{\ell^2+\ell -2}{2r^2}\right\}\\
   &+\frac{2m h_0}{raf} \left\{-\frac{r a''-3 a' m'}{r a}+\frac{m }{r^2 a}\left(3 a'-2r a''+ r \zeta\right)\right.\\
   &\left.-f|\psi '|^2+\frac{  m''}{r}-\mathcal{M}_{\rm s}|\psi|^2-\frac{\ell ^2-\ell}{2r^2}+\frac{2m }{r^3 } \right\} \\
   &+\dot h_0\left(-\frac{4m^2 a'}{a^3f r^2}+\frac{2 r m\left(m'-1\right)-2 m^2+r^2}{
   a^2 r^3 f^2}\right)\\
   &+\frac{m \left(r a'-2 a\right) \dot h_1}{r^2 a^2}-\frac{m a' h'_0}{r a^2}-\frac{\left(r^2-8
   m^2\right) \dot h'_0}{2 r^2 a^2 f}+\frac{m h''_0}{r a}\\
   &-\frac{m \dot h'_1}{r a}-\frac{m \gamma^2
   \ddot h_0}{r a^3 f}+\frac{\gamma^2 \ddot h_1}{2 a^2}=0\,,
\end{aligned}
\label{eq:ax:23}
\end{equation}
where overhead dots refer to time derivatives. Similar to before, it is convenient to introduce a time-domain master function
\begin{equation}
\mathcal{Q}(t,r)=\frac{q(r)}{r}h_1(t,r)\,,
\end{equation}
with
\begin{equation}
   q(r)= c_1 a f \sqrt{m}\,.\label{eq:ax:q} 
\end{equation}
The constant $c_1$ is an arbitrary integration constant, and we set $c_1=\sqrt{2/r_{\rm h}}$ to recover the Regge-Wheeler function in the vacuum limit. Using the definitions above one obtains the hyperbolic second-order equation used in the time evolutions [cf. Eq.~\eqref{eq:Qcal}], with
\begin{equation}
\begin{aligned}
    {\mathcal B}(r)&=-a^2f^2 \frac{m'}{m}\,, \\
    \mathcal{V}(r)&=\frac{a f m' \left(r (r-14 m) a'+6 a m\right)}{2 r^2 m}\\
   &
    +\frac{a f \left(r \left(2 m \left(2 a'+\zeta  r\right)+r
   a'\right)+a (r \ell  (\ell +1)-6 m)\right)}{r^3}\\
   &+2 a f^2 a''-\frac{a^2 f (3 r-2 m) m'^2}{4 r m^2}+\frac{a^2
   f (r-6 m) m''}{2 r m}\\
   &-2 a^2 f |\psi|^2 \left(\frac{2 |\Omega|^2 m(r)}{ra^2}-\mathcal{M}_s\right)+2 a^2 f^2 |\psi
   '|^2\,.
\end{aligned}
\end{equation}
\bibliography{references}
\end{document}